\input{aipcheck}

\documentclass[
    ,final            
  ]
  {aipproc}

\layoutstyle{6x9}

\begin{document}

\title{Spitzer Surveys of IR Excesses of White Dwarfs}



\author{You-Hua Chu}{
  address={Astronomy Department, University of Illinois}
}

\author{Robert A. Gruendl}{
  address={Astronomy Department, University of Illinois}
}

\author{Jana Bilikova}{
  address={Astronomy Department, University of Illinois}
}

\author{Andrew Riddle}{
  address={Astronomy Department, University of Illinois}
}

\author{Kate Y.-L. Su}{
  address={Steward Observatory, University of Arizona}
}

\begin{abstract}
IR excesses of white dwarfs (WDs) can be used to diagnose the 
presence of low-mass companions, planets, and circumstellar dust.
Using different combinations of wavelengths and WD temperatures, 
circumstellar dust at different radial distances can be surveyed.
The {\it Spitzer Space Telescope} has been used to search for
IR excesses of white dwarfs.  Two types of circumstellar dust disks
have been found: (1) small disks around cool WDs with 
$T_{\rm eff} < 20,000$ K, and (1) large disks around hot WDs with
$T_{\rm eff} > 100,000$ K.  The small dust disks are within the Roche
limit, and are commonly accepted to have originated from tidally 
crushed asteroids.  The large dust disks, at tens of AU from the 
central WDs, have been suggested to be produced by increased
collisions among Kuiper Belt-like objects.  In this paper, we 
discuss {\it Spitzer} IRAC surveys of small dust disks around
cool WDs, a MIPS survey of large dust disks around hot WDs, and
an archival {\it Spitzer} survey of IR excesses of WDs.
\end{abstract}

\maketitle


\section{Introduction}
The photospheric emission of white dwarfs (WDs) has relatively 
simple spectra and can be approximated by the Rayleigh-Jeans Law
at infrared (IR) wavelengths.  If a WD shows much higher IR
fluxes than expected from its photosphere, there must be an
external object contributing to the IR emission, such as
a low-mass stellar or sub-stellar companion, a planet, or
a circumstellar dust disk.
Near-IR excesses in the $JHK$ bands have been routinely used to
search for low-mass companions.  Observations at longer
wavelengths have been successful in finding brown dwarf
companions and dust disks around WDs, but no planets around
WDs have been unambiguously detected through IR excesses.

The launch of the {\it Spitzer Space Telescope} made it
possible to perform imaging surveys in the 3.6, 4.5, 5.8, and 
8.0 $\mu$m bands with the Infrared Array Camera 
\citep[IRAC,][]{Fetal04}, and in
the 24, 70, and 160 $\mu$m bands with the Multiband Imaging 
Photometer for {\it Spitzer} \citep[MIPS,][]{Retal04}.  
In addition, the Infrared Spectrograph \citep[IRS,][]{Hetal04}
on-board {\it Spitzer} can be used to take spectra from 5 to 
37 $\mu$m.  These instruments are ideal for investigating the
frequency of occurrence and physical nature of IR excesses of WDs.

We are mainly interested in IR excesses caused by circumstellar
dust disks of WDs.  Debris disks of stars should have dissipated
well before the stars evolve off the main sequence
\citep[e.g.][]{Retal05}.  The circumstellar dust of WDs must
have been generated recently.  G29-38 and GD362 are the first 
two WDs reported to possess circumstellar dust disks 
\citep{Betal05,Ketal05,Retal05,ZB87}.  These dust disks have been 
suggested to originate from tidally crushed asteroids, and this
origin is supported by their small distances from the central WD 
(within the Roche limit), silicate features in the dust 
emission, and high metal content in the central WD's atmosphere
\citep{Jura03,Jetal07}.
A much larger dust disk has been discovered around the
central WD of the Helix planetary nebula (PN); this dust
disk, extending 35-150 AU from the central WD, is suggested 
to be produced by collisions among Kuiper Belt-like objects
\citep{Setal07}.

These two types of dust disks can be understood in the context of 
stellar evolution.  As an intermediate- or low-mass star evolves
to the final WD stage, a large fraction of its initial mass is
lost to form a PN.  If a star has a planetary system, the 
reduced stellar gravitational field will cause the
planetary system to expand.  The expansion rejuvenates the 
internal kinematics of the planetary system.
Through scattering and orbital resonances with giant planets, the
sub-planetary objects may experience higher collision rates and
produce dust along their orbits, and some sub-planetary objects
may be scattered into the Roche limit and be tidally
pulverized.  

The solar system has an Asteroid Belt at $\sim$3 AU
and a Kuiper Belt at 30-50 AU from the Sun.  
Using the solar system analog, we expect dust around WDs to
be produced within (1) the Roche limit, (2) the Asteroid Belt,
and (3) the Kuiper Belt.  Considering that the progenitors of 
WDs have a range of masses and that planetary systems around 
WDs have expanded from their original locations, we expect these
dust disks around WDs to be detected at radial distances of 
(1) $<$0.01 AU, (2) a few to over 10 AU, and (3) a few tens 
to over 100 AU, respectively.

This paper will focus on {\it Spitzer} surveys for dust disks 
around WDs, how they target these different types of dust disks.

\section{Spitzer Wavebands and Expected Dust Emission}

The {\it Spitzer Space Telescope} can be used to image dust
emission in the 3.6, 4.5, 5.8, 8.0, 24, 70, and 160 $\mu$m bands
in principle, but the poorer sensitivity and angular resolution of
MIPS at 70 and 160 $\mu$m make it less desirable to survey in
these two bands.  We will thus examine the feasibility of
using the IRAC bands and the MIPS 24 $\mu$m band to survey
circumstellar dust disks of WDs.

The equilibrium temperature of a dust grain can be calculated
by equating the absorbed stellar radiation to the radiated
dust emission.  Assuming a total absorption (i.e. zero albedo)
and an earth radius ($R_\oplus$) for the WD, the dust temperature 
($T_{\rm g}$)
can be approximated by $T_{\rm g}$ = $T_{\rm WD}(R_\oplus/2 D)^{1/2}$,
where $T_{\rm WD}$ is the effective temperature of the WD, and 
$D$ is the radial distance from the WD.
Figure 1 plots the curves of dust temperatures as a function
of $T_{\rm WD}$ and $D$.  The dust temperatures are selected 
so that the peak emission wavelengths match the {\it Spitzer}
wavebands.

The top panel of Figure 1 is optimized to illustrate dust
temperatures for tidally crushed asteroids.  The highest
dust temperature plotted is 1500 K, the sublimation temperature.
The Roche limit is plotted in solid and dash-dot lines for 
asteroids of densities 1 and 3 g~cm$^{-3}$, respectively.
The shaded area shows the range of stellar temperatures
for which the dust emission from tidally crushed asteroids 
can be detected in the {\it Spitzer} bands.  It is clear that
{\it only WDs with effective temperatures lower than $\sim$20,000 K
can possess such dust disks.}  For WDs with higher temperatures,
the crushed asteroids may sublimate and form gaseous disks,
such as those reported by \citet{Getal07}. The {\it Spitzer} 
IRAC bands are ideal for surveys of such small dust disks.

\begin{figure}[!t]
\resizebox{0.5\columnwidth}{!}
 {\includegraphics{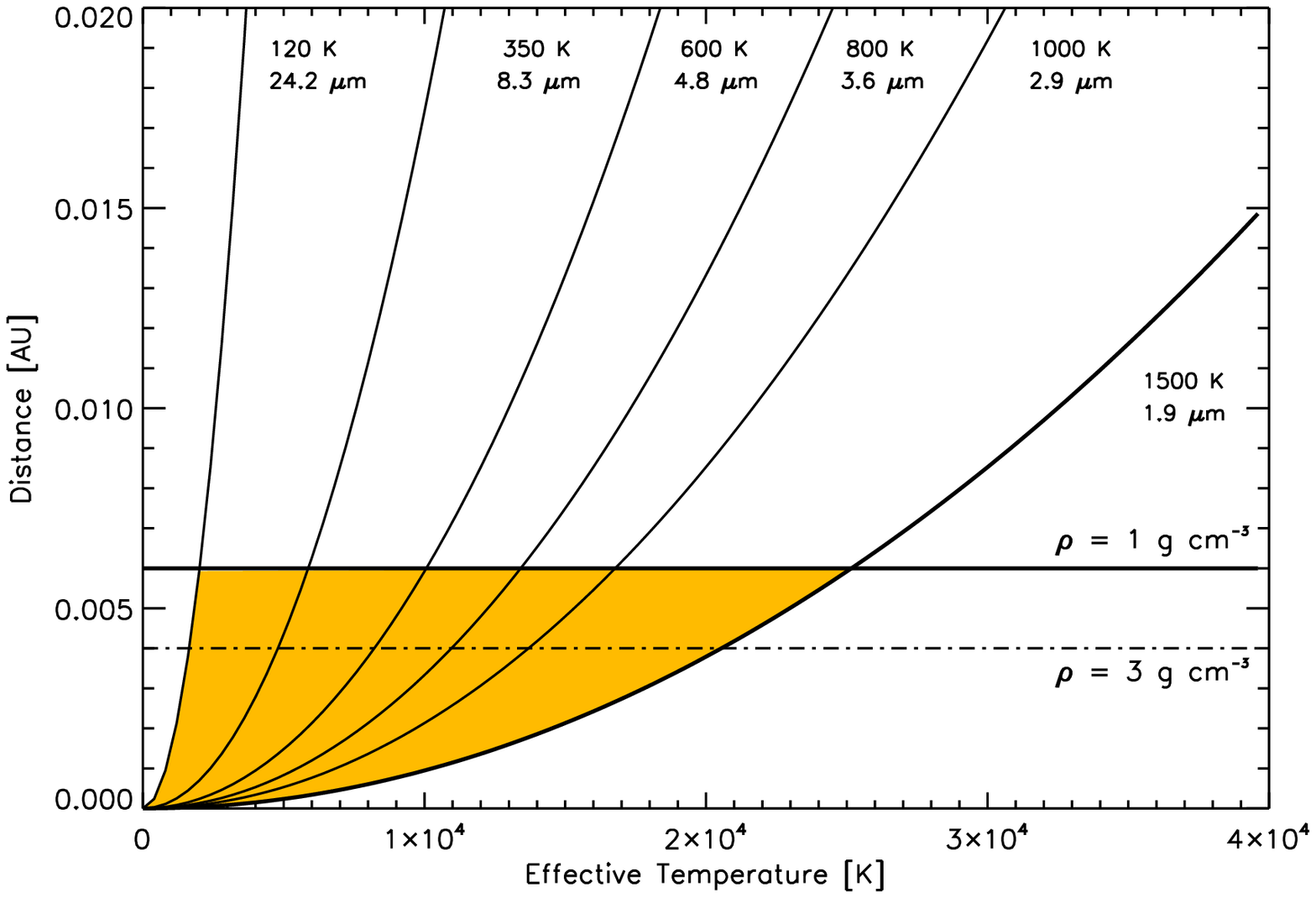}}
\end{figure}
\begin{figure}[!t]
\resizebox{0.5\columnwidth}{!}
 {\includegraphics{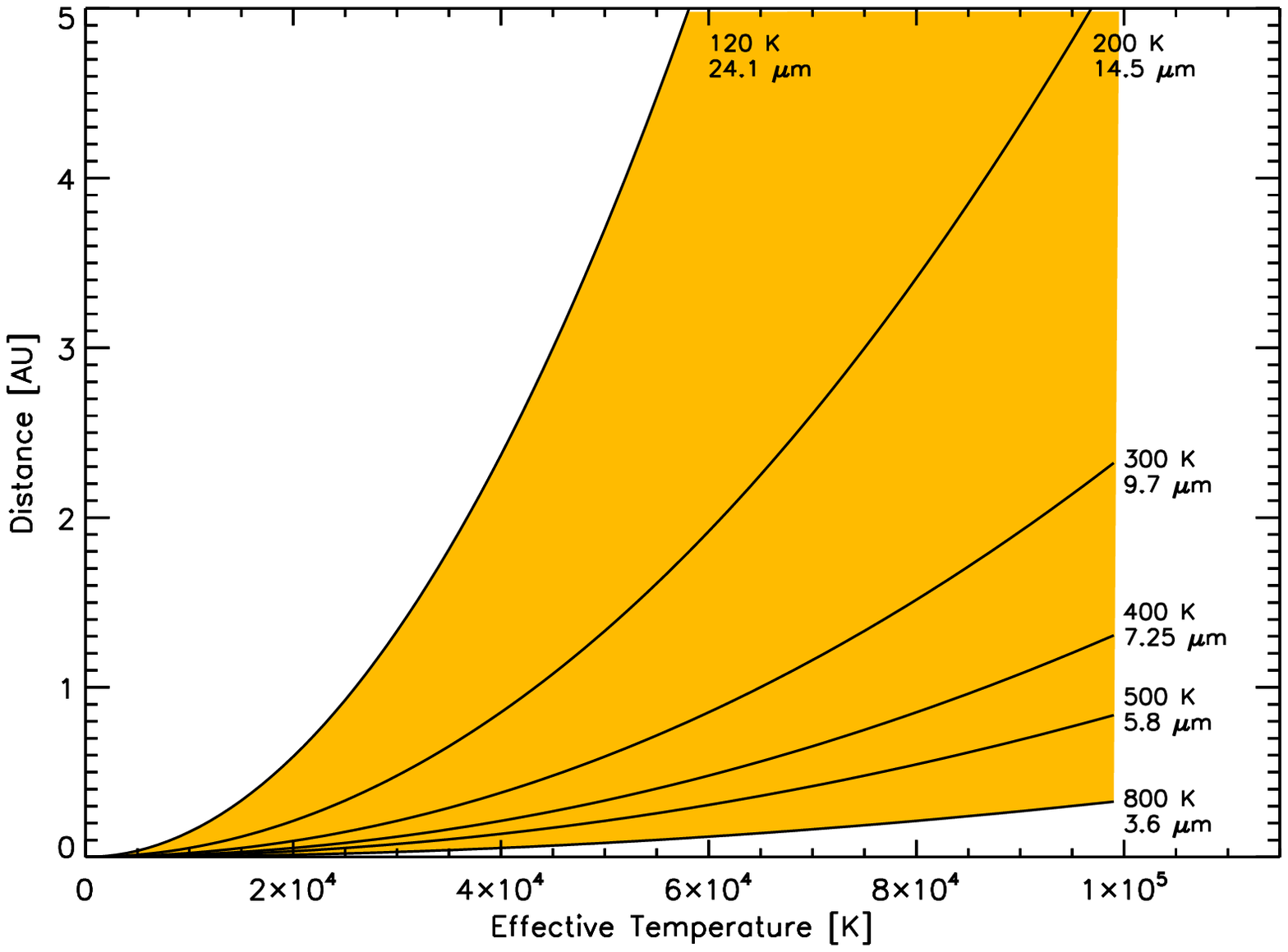}}
\resizebox{0.5\columnwidth}{!}
 {\includegraphics{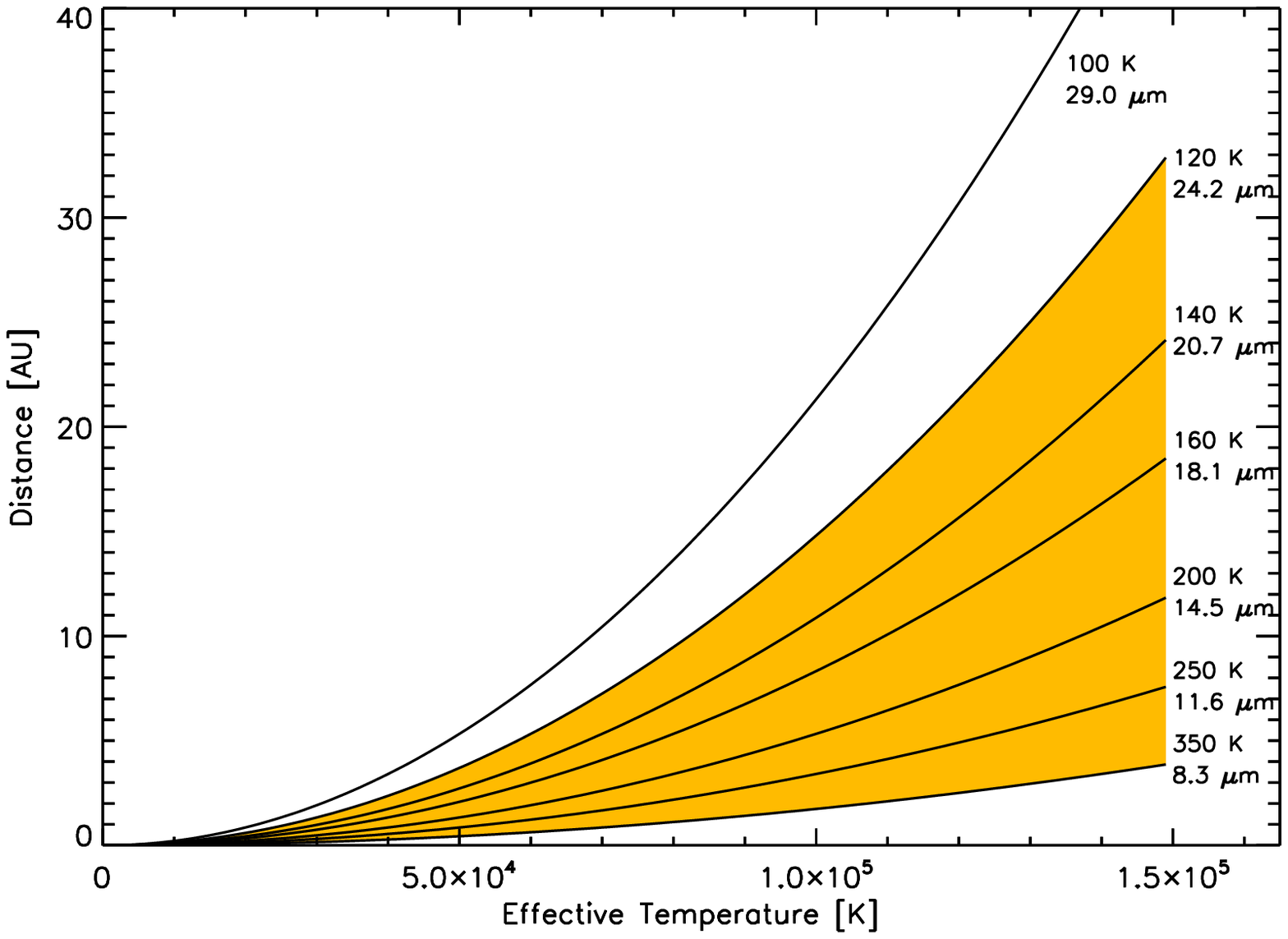}}
\caption{Iso-temperature curves of dust as a function of
stellar effective temperature and radial distance to the
central star.  Some temperatures are selected so that the 
blackbody peak emission wavelengths are comparable to the 
central wavelengths of {\it Spitzer} IRAC bands and MIPS 
24~$\mu$m band.  The three panels are optimized to show dust
temperatures for three different ranges of radial distances.
In the top panel, the dust sublimation temperature (1500 K)
curve is plotted for reference, and the Roche limits for 
asteroids of densities 1 and 3 g~cm$^{-3}$ are plotted in 
solid and dash-dot lines, respectively.  The shaded area 
marks the parameter space where dust emission from tidally 
crushed asteroids can be detected.  The two lower panels
show the expected dust temperatures at distances of the
Asteroid Belt and Kuiper Belt, respectively.}
\label{fig:1}
\end{figure}

The lower left panel of Figure 1 shows that at 3-5 AU
from a WD, roughly where the Asteroid Belt is 
from the sun,
the dust temperature is so low that the peak emission
wavelength is much longer than those of the IRAC bands. 
{\it MIPS 24 $\mu$m surveys of WDs $\sim$50,000 K may 
detect dust emission at radial distances of a few AU.}
Emission from dust in Asteroid Belts can be surveyed at 
longer or shorter wavelengths for WDs of lower or higher 
temperatures, respectively, but not with {\it Spitzer}.

The lower right panel of Figure 1 shows that at a few
tens of AU from a WD, roughly where the Kuiper Belt is
located, the dust temperature is still lower.  {\it Only
MIPS 24 $\mu$m surveys for WDs with effective temperatures
$\sim$100,000 K or hotter may detect dust emission at
radial distances of a few tens of AU.}  For any cooler WDs,
surveys for such cold dust need to be carried out at 
longer wavelengths.

\section{{\it Spitzer} Surveys of White Dwarfs}

Many {\it Spitzer} programs have been designed to survey
WDs. Some are general surveys, and some target specific
types of WDs.  The entire {\it Spitzer} archive contains
many observations that targeted other objects but 
serendipitously included WDs in the fields.  Roughly
four categories of {\it Spitzer} surveys of WDs exist:

\begin{itemize}

\item Broad survey programs by Kuchner (P2313; IRAC), Kilic
(P474; IRAC), Farihi (P30807; IRAC), and Chu (P40953; MIPS).
A total of $\sim$310 WDs were surveyed in these programs.

\item Survey programs targeting metal-rich WDs, such as
DZ, DAZ, DBZ, etc., by Jura (P275 and P50060; IRAC), Burleigh
(P30432 and P60161; IRAC), Farihi (P30807, P60119, P70037, and
P70116; IRAC), Debes (3655; IRAC), and Kilic (P70023; IRAC).
A total of about 80--90 WDs were observed in these programs.

\item Survey for WDs originating from binary mergers by
Hansen (P3309; IRAC).  Fourteen WDs were observed in this
program.

\item {\it Spitzer} archive contains serendipitous IRAC
observations for $\sim$330 WDs; among these $\sim$130 WDs are
detected.

\end{itemize}

We can examine the temperatures of WDs observed by 
{\it Spitzer} to see what types of dust disks may be detected.
Figure 2 shows histograms of temperature distributions for
WDs targeted specifically in {\it Spitzer} IRAC and
MIPS observations, and those that were detected 
serendipitously in archival data.
The top panel shows that {\it Spitzer} IRAC observations
were mostly made for cooler WDs.  For WDs cooler than 
20,000 K, IRAC observations may detect dust emission from
tidally disrupted asteroids. IRAC observations of WDs
hotter than 20,000 K may detect dust emission at
radii of $\sim$1 AU of higher, but no IR excesses in IRAC
bands are detected for WDs hotter than 20,000 K.

The middle panel of Figure 2 shows the temperature 
distribution of WDs targeted by MIPS 24 $\mu$m observations.
Two distinct groups of WDs were observed.  In the cooler
group, most of the WDs were observed to search for outer
boundaries of dust disks produced by tidally crushed 
asteroids.  In the hotter group, most of the WDs were 
specifically targeted to search for dust in the Kuiper 
Belt-like distances.

The bottom panel of Figure 2 shows the temperature 
distribution of WDs serendipitously detected in archival
{\it Spitzer} IRAC observations.  Preliminary
examination of their optical-to-IR spectral energy 
distributions (SEDs) suggests that $\sim$10\% show 
IR excesses due to late-type companions or 
circumstellar dust.  We are still analyzing these objects in 
greater detail to determine the origin of the IR excesses.

Of all {\it Spizter} surveys of WDs, the most successful 
in finding circumstellar dust disks are the IRAC surveys
of cool, metal-rich WDs \citep[e.g.][]{Fetal10,Jetal07}
and the MIPS survey of hot WDs.
As we now understand that the dust created by tidally
destroyed asteroids will be accreted onto the central
WD and enrich its atmosphere, it is thus not surprising
that IRAC surveys of metal-rich WDs have much higher 
success rates than broad IRAC surveys of WDs 
\citep[e.g.][]{Metal07}.
The MIPS survey of hot WDs, discussed in the next section,
also detected IR excesses for a significant fraction of the
targets.  These would correspond to dust emission at distances
of a few tens to 100 AU, equivalent to the Kuiper Belt, from the 
central WD.  It is worth noting that no dust disks around
WDs at radial distances of a few AU have been detected. 
This is because such dust disks can be detected around
WDs of temperatures $\sim$50,000 K using MIPS 24 $\mu$m
observations (lower left panel of Figure 1), but hardly any 
such observations were made during the cryogenic mission of 
{\it Spitzer} (middle panel of Figure 2).

\begin{figure}[!t]
\resizebox{0.67\columnwidth}{!}
 {\includegraphics{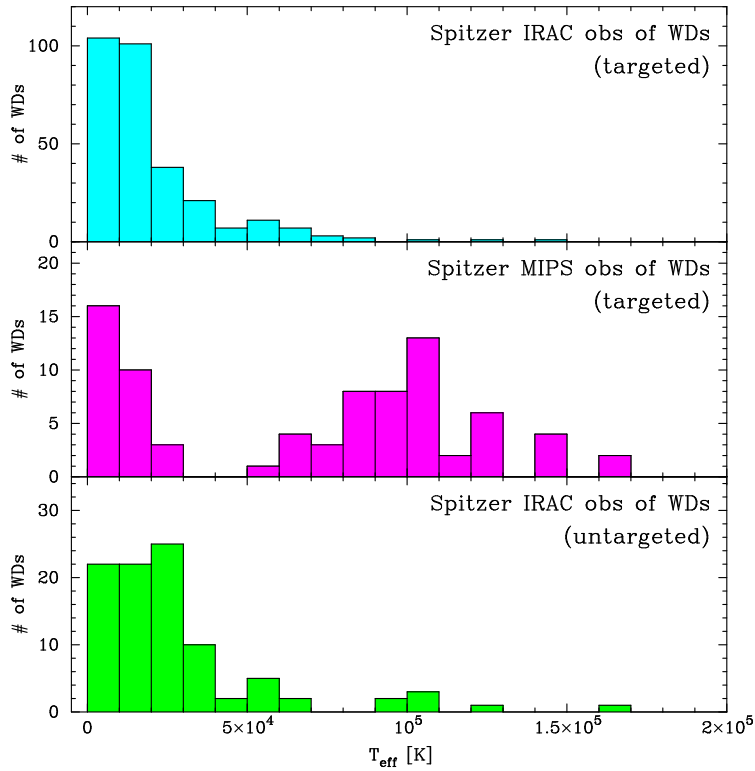}}
\caption{Temperature distributions of WDs observed
by {\it Spitzer} in targeted IRAC and MIPS observations
and WDs serendipitously detected in archival
{\it Spitzer} IRAC observations.}
\end{figure}

\section{{\it Spitzer} MIPS 24 $\mu$m Survey of Hot White Dwarfs}

{\it Spitzer} observations of the Helix Nebula 
revealed a bright, compact source at the central
WD in the MIPS 24 and 70 $\mu$m bands.  Follow-up 
{\it Spitzer} IRS spectra of this WD confirm that
the IR emission is continuum in nature and consistent
with a blackbody temperature of 90-130 K.  The large
emitting area, 3.8--38 AU$^2$, can be provided only
by a dust cloud.  As the WD does not suffer a large
extinction, the dust is likely distributed in a disk,
and the disk extends 35 to 150 AU from the central
WD, of which the stellar effective temperature is
110,000 K.  As the spatial extent of this dust disk
corresponds to that of the Kuiper Belt in the solar
system, it is suggested that the dust disk around the
central WD of the Helix Nebula was produced by 
collisions among Kuiper Belt-like objects \citep{Setal07}.

Inspired by the central WD of Helix, we have carried out a
MIPS 24 $\mu$m survey of 71 hot WDs or pre-WDs in
evolved PNe \citep{Cetal10}. 
The majority of the targets have effective temperatures
greater than 100,000 K.
Nine of these targets were detected in the 24 $\mu$m 
observations and each represents an IR excess.
These nine objects are listed in Table 1, where the
Helix Nebula is also included for comparison.
Follow-up {\it Spitzer} IRS spectra of K1-22, WD\,0103+732,
WD\,0127+581, WD\,0439+466, and WD\,0950+139 show that in
every case the emission in the 24 $\mu$m band is dominated by 
dust continuum.  The SEDs and spectral analysis are presented
in the paper by Bilikova et al.\ in this volume.

\begin{table}[t!]
\begin{tabular}{lllccc}
\hline
WD Name     &    PN Name   & WD Type &$T_{\rm eff}$ (kK)& $F_{24}$ (mJy) & $L_{\rm IR}/L_*$ \\
\hline
K1-22       &    K1-22     &  ...    &    141       &   ~1.07   & $3.1\times10^{-5}$ \\
NGC 2438    &    NGC 2438  &  ...    &    114       &   12.4~   & $4.5\times10^{-4}$ \\
WD 0103+732 &    EGB 1     &  DA     &    150       &   ~2.76   & $1.3\times10^{-5}$ \\
WD 0109+111 &    ...       &  DOZ    &    110       &   ~0.27   & $4.9\times10^{-6}$ \\
WD 0127+581 &    Sh2-188   &  DAO    &    102       &   ~0.34   & $2.7\times10^{-5}$ \\
WD 0439+466 &    Sh2-216   &  DA     &    ~95       &   ~0.98   & $3.7\times10^{-6}$ \\
WD 0726+133 &    Abell 21  &  PG1159 &    130       &   ~0.92   & $1.6\times10^{-5}$ \\
WD 0950+139 &    EGB 6     &  DA     &    110       &   11.7~   & $2.6\times10^{-4}$ \\
WD 1342+443 &    ...       &  DA     &    ~79       &   ~0.22   & $4.0\times10^{-5}$ \\
\hline
WD 2226-210 &    Helix Neb &  DAO    &    110       &   48.0~   & $2.5\times10^{-4}$\\
\hline
\end{tabular}
\caption{{\it Spitzer} MIPS 24 $\mu$m Detection of Hot WDs and Pre-WDs}
\end{table}

The summary in Table 1 shows that a great majority of the hot WDs
exhibiting 24 $\mu$m excesses are still in PNe, and the two faintest
24 $\mu$m sources belong to WDs without visible PNe.  
Unlike the WDs with dust disks from tidally crushed asteroids, the
WDs with 24 $\mu$m excesses do not necessarily show metal-enriched
atmospheres, indicating that the dust accretion rates must be low.

Nine of the 71 WDs surveyed show 24 $\mu$m excesses.  The apparent 
detection rate is 12-13\%.  We find that the WD targets without 
2MASS $J$ band measurements (due to their faintness) have a lower 
detection rate than the brighter WD targets with 2MASS $J$ magnitudes.
This is presumably a bias introduced by distance - distant WDs are 
fainter and  detected at optical wavelengths but not in $J$ by 2MASS.
Using the brighter subsample of hot WDs,
we conclude that at least 15\% hot WDs show 24 $\mu$m excesses.






\bibliographystyle{aipproc}   

\bibliography{sample}


\end{document}